\documentstyle[epsf,12pt]{article}
\input{epsf}
\input{psfig}
\newlength{\dinwidth}
\newlength{\dinmargin}
\begin{document}
\newcommand{\Gevsq}     {\mbox{${\rm GeV}^2$}}
\newcommand{\Kt}        {\mbox{$k_\perp$}}
\newcommand{\Pt}        {\mbox{$p_T$}}
\newcommand{\qsd}       {\mbox{${Q^2}$}}
\newcommand{\yjb}       {\mbox{${y_{_{JB}}}$}}
\newcommand{\xda}       {\mbox{$x_{_{DA}}$}}
\newcommand{\qda}       {\mbox{$Q^{2}_{_{DA}}$}}
\newcommand{\gsubh}     {\mbox{$\gamma_{_{H}}$}}
\newcommand{\thee}      {\mbox{$\theta_{e}$}}
\newcommand{\mrsdm}     {\mbox{MRSD$_-^{\prime}$\ }}
\newcommand{\mrsdz}     {\mbox{MRSD$_0^{\prime}$\ }}
\newcommand{\sleq} {\raisebox{-.6ex}{${\textstyle\stackrel{<}{\sim}}$}}
\newcommand{\sgeq} {\raisebox{-.6ex}{${\textstyle\stackrel{>}{\sim}}$}}
\renewcommand{\thefootnote}{\arabic{footnote}}
\title{
{\bf Study of the Photon Remnant in Resolved Photoproduction at HERA} \\
\author{\rm ZEUS Collaboration}
}
\date{ }
\maketitle
\vspace{5 cm}
\begin{abstract}
Photoproduction at HERA is studied in $ep$ collisions, with the ZEUS detector,
for
$\gamma p$ centre-of-mass energies ranging from 130-270 GeV. A sample of events
with two high-\Pt\ jets (\Pt\ $> 6$ GeV, $\eta <1.6$) and a third cluster in
the
approximate direction of the electron beam is isolated using a clustering
algorithm.  These events are mostly due to resolved photoproduction.  The third
cluster is identified as the photon remnant. Its properties, such as the
transverse and longitudinal energy flows around the axis of the cluster, are
consistent with those commonly attributed to jets, and in particular with those
found for the two jets in these events.  The mean value of the photon remnant
\Pt\
with respect to the beam axis is measured to be $2.1 \pm 0.2$ GeV, which
demonstrates substantial mean transverse momenta for the photon remnant.
\end{abstract}
\vspace{-18cm}
\begin{flushleft}
\tt DESY 95-083 \\
April 1995 \\
\end{flushleft}

\setcounter{page}{0}
\thispagestyle{empty}

\newpage
%

%   30/03/95 504231752  MEMBER NAME  AUTH023  (ZEUS)     M  TEX
%
\def\3{\ss}
\parindent 0cm
\footnotesize
\renewcommand{\thepage}{\Roman{page}}
\begin{center}
\begin{large}
The ZEUS Collaboration
\end{large}
\end{center}
M.~Derrick, D.~Krakauer, S.~Magill, D.~Mikunas, B.~Musgrave,
J.~Repond, R.~Stanek, R.L.~Talaga, H.~Zhang \\
{\it Argonne National Laboratory, Argonne, IL, USA}~$^{p}$\\[6pt]
R.~Ayad$^1$, G.~Bari, M.~Basile,
L.~Bellagamba, D.~Boscherini, A.~Bruni, G.~Bruni, P.~Bruni, G.~Cara
Romeo, G.~Castellini$^{2}$, M.~Chiarini,
L.~Cifarelli$^{3}$, F.~Cindolo, A.~Contin, M.~Corradi,
I.~Gialas$^{4}$,
P.~Giusti, G.~Iacobucci, G.~Laurenti, G.~Levi, A.~Margotti,
T.~Massam, R.~Nania, C.~Nemoz, \\
F.~Palmonari, A.~Polini, G.~Sartorelli, R.~Timellini, Y.~Zamora
Garcia$^{1}$,
A.~Zichichi \\
{\it University and INFN Bologna, Bologna, Italy}~$^{f}$
% ...
\\[6pt]
\clearpage
\hspace*{1mm}
$^{ 1}$ supported by Worldlab, Lausanne, Switzerland \\
\hspace*{1mm}
$^{ 2}$ also at IROE Florence, Italy  \\
\hspace*{1mm}
$^{ 3}$ now at Univ. of Salerno and INFN Napoli, Italy  \\
\hspace*{1mm}
$^{ 4}$ supported by EU HCM contract ERB-CHRX-CT93-0376 \\
\hspace*{1mm}
% ...
$^{18}$ now with OPAL Collaboration, Faculty of Physics at Univ. of
        Freiburg \\
% ...

\begin{tabular}{lp{15cm}}
$^{a}$ &supported by the Natural Sciences and Engineering Research
         Council of Canada (NSERC) \\
$^{b}$ &supported by the FCAR of Qu\'ebec, Canada\\
$^{c}$ &supported by the German Federal Ministry for Research and
         Technology (BMFT)\\
$^{d}$ &supported by the MINERVA Gesellschaft f\"ur Forschung GmbH,
         and by the Israel Academy of Science \\
$^{e}$ &supported by the German Israeli Foundation, and
         by the Israel Academy of Science \\
$^{f}$ &supported by the Italian National Institute for Nuclear Physics
         (INFN) \\
$^{g}$ &supported by the Japanese Ministry of Education, Science and
         Culture (the Monbusho)
         and its grants for Scientific Research\\
$^{h}$ &supported by the Korean Ministry of Education and Korea Science
         and Engineering Foundation \\
$^{i}$ &supported by the Netherlands Foundation for Research on Matter
         (FOM)\\
$^{j}$ &supported by the Polish State Committee for Scientific Research
         (grant No. SPB/P3/202/93) and the Foundation for Polish-
         German Collaboration (proj. No. 506/92) \\
$^{k}$ &supported by the Polish State Committee for Scientific
         Research (grant No. PB 861/2/91 and No. 2 2372 9102,
         grant No. PB 2 2376 9102 and No. PB 2 0092 9101) \\
$^{l}$ &partially supported by the German Federal Ministry for
         Research and Technology (BMFT) \\
$^{m}$ &supported by the German Federal Ministry for Research and
         Technology (BMFT), the Volkswagen Foundation, and the Deutsche
         Forschungsgemeinschaft \\
$^{n}$ &supported by the Spanish Ministry of Education and Science
         through funds provided by CICYT \\
$^{o}$ &supported by the Particle Physics and Astronomy Research
        Council \\
$^{p}$ &supported by the US Department of Energy \\
$^{q}$ &supported by the US National Science Foundation
\end{tabular}

\newpage
\pagenumbering{arabic}
\setcounter{page}{1}
\normalsize

\section{\bf Introduction}

Hard scattering between a real photon and a proton is expected,
in lowest order QCD, to occur
by two different mechanisms (Fig.\ 1).  The photon may interact
directly with a
quark or gluon from the proton or it may resolve
into its constituent quarks and gluons which then interact
with the partons from the proton.  These two processes are called direct and
resolved photoproduction, respectively \cite{theory}.

In direct photoproduction, the entire photon participates
in the hard scatter.
In resolved photoproduction on the other hand,
only a fraction of the momentum of the photon
is involved in the hard scatter while the
remaining momentum is carried by spectator partons.
These partons fragment into a photon remnant which is
expected to be approximately collinear with
the original photon.
The presence of resolved photon interactions at HERA has been
demonstrated \cite{res-ZE,res-H1} and the existence
of the photon remnant has been confirmed.  The separation
between the direct and resolved contributions has also been reported,
together with the measurement of a differential dijet cross section~\cite{xg}.
Inclusive jet \cite{jet-h1,jet-ze} and dijet \cite{dijet-ze} cross sections
have given further information on the kinematic properties of
these two processes
and on the gluon content of the proton as well
as on the parton structure of the photon.

The presence of a photon remnant in photon-proton collisions
has been used as a means to identify
resolved photoproduction interactions.
The study of the photon remnant itself, however, is
also interesting: a detailed
comparison with leading-order (LO) predictions
has not been performed and little is known about the
internal structure of the photon remnant.  On the theoretical side,
although the point-like coupling of
the photon to quark-antiquark pairs is included in parameterizations
of the photon structure function, most Monte Carlo simulations model the
resolved photon as a hadron, with collinear incoming partons.
This results in a photon remnant with low-\Pt\ with respect to the
beam axis.
Several studies have suggested that next-to-leading-order contributions
or fluctuations of the photon into quark-antiquark pairs with
high virtuality
may lead to a `photon remnant' which has sizable transverse momentum
with respect to the incident photon direction
\cite{Sjostrand,Chyla,drees}.

In this paper, the photon remnant is isolated for the first time.
Its properties are studied using a clustering algorithm and are
found to be consistent with those commonly attributed to jets.
These remnant jets are then compared with the jets emerging from
the hard interaction and with LO Monte Carlo expectations.
The data were collected with the ZEUS detector during the 1993
data-taking period.
The study is conducted for $\gamma p$ centre-of-mass energies
($W_{\gamma p}$) in the range 130 $< W_{\gamma p} <$ 270 GeV.

\section{\bf Experimental setup }

Details of the ZEUS detector have been described elsewhere~\cite{ZEUS}.
The primary components used in this analysis are the calorimeter and the
tracking detectors. The
uranium-scintillator calorimeter \cite{CAL} covers 99.7\% of the
total solid angle.  It is subdivided into electromagnetic and hadronic
sections with cell sizes of $5 \times 20$ cm$^2$ ($10
\times 20$ cm$^2$
in the outgoing electron direction\footnote{The $Z$ axis is defined
to lie along the proton
direction; the $Y$ axis points upward; the pseudorapidity,
$\eta = - $ln$(\tan\frac{\theta}{2}$) where $\theta$ is
the angle with respect to the $Z$ axis.})
and $20 \times 20$ cm$^2$, respectively.
It consists of three parts: the rear calorimeter (RCAL) covering the
region $-3.8 < \eta < -0.75$,
the barrel calorimeter (BCAL) covering the region $-0.75 < \eta < 1.1$
and the forward calorimeter (FCAL) covering the region
1.1 $ < \eta < $ 4.3.
The calorimeter has an energy resolution achieved in test beams
of $\sigma/E =
18 (35)$ \%/$\sqrt{E ({\rm GeV})}$ for electrons (hadrons).
The timing resolution for each cell is
$\sigma_t$ =1.5/$\sqrt E$ $\oplus$ 0.5 ns, where $E$ (GeV) is the energy
deposited in the cell.

The tracking system consists of a vertex detector \cite{VXD}
and a central tracking chamber \cite{CTD} inside a 1.43 T solenoidal
magnetic field.  The interaction vertex
is measured with a resolution along (transverse to) the beam direction
of 0.4 (0.1) cm.

The luminosity is measured, using the electron-proton
bremsstrahlung process, by electron and photon
lead-scintillator calorimeters \cite{LUMI} installed inside the HERA tunnel.
In 1993 the beam energies at HERA were $E_e = 26.7$ GeV for the electrons
and $E_p = 820$ GeV for the protons.  Typical electron
and proton currents were about 10 mA and the instantaneous luminosity
was about 6 $\times$ 10$^{29}$ cm$^{-2}$ s$^{-1}$.
HERA operated with 84 colliding bunches.
Additional electron and proton bunches circulated without colliding
and are used for background measurements.

\section{\bf Trigger and data selection}

In this analysis, photoproduction events are defined by requiring
that the electron was scattered at small angles and was not detected
in the calorimeter.  This requirement corresponds approximately to a cut
of $Q^2 \le 4$ GeV$^2$, giving a median $Q^2$ of about $10^{-3}$
GeV$^2$~\cite{xg}.
The trigger selects hard scattering events at low $Q^2$.

The ZEUS detector uses a three level trigger \cite{ZEUS}.
In the first level trigger,
the calorimeter cell energies were combined to define
regional and global sums which were required to exceed given thresholds
\cite{wsmith}.
The second level trigger mainly rejected
beam-gas interactions using timing information from the
calorimeter. The third level trigger performed further
rejection of
beam-gas and cosmic ray events using information from both the calorimeter and
the tracking chambers.  An event was rejected if no vertex was
found by the central tracking chambers or if the vertex was located
in the region $|Z| > 75$ cm.
To reject beam-gas interactions, events were
selected based on the following kinematic
cuts~\cite{jet-ze}:
$E_{tot} - p_Z \geq$ 8 GeV,
$p_Z/E_{tot} \leq$ 0.94 and
$E_T^{cone} \geq$ 12 GeV, where the calorimetric quantities
$E_{tot}$, $p_Z$ and $E_T^{cone}$ are
the total energy, the total longitudinal energy
and the transverse energy
excluding a cone of 10$^\circ$ in the forward direction, respectively.
About 470,000 triggers were collected with these trigger conditions.

As in previous studies of hard photoproduction
\cite{jet-ze,dijet-ze}, the following offline cuts were applied
to select the final event sample.
\begin {itemize}
\item Beam-gas interactions were reduced by tightening calorimeter
timing cuts, as well as cuts on the correlation between
the vertex position (defined by the tracking chambers) and
the calorimeter timing \cite{DIS1}.

\item The $E_T^{cone}$ cut was raised to $E_T^{cone} \ge $ 15 GeV to select
hard scattering events.

\item To reduce beam-gas interactions, the event was rejected
if less than 10\% of the tracks pointed toward the vertex.

\item Deep inelastic scattering (DIS) neutral current events were removed
from the sample as described in our previous publications \cite{res-ZE,xg}.

\item The fraction of the initial
electron energy carried by the almost real photon, $y =
E_{\gamma}/E_e$ where $E_{\gamma}$ is the photon energy, was
measured using the Jacquet-Blondel~\cite{yjb} estimator of the Bjorken-$y$:
\[
y_{_{JB}} = \frac{\Sigma_{i}(E^{i}-p^{i}_{Z})}{2E_{e}}.
\]
The sum runs over calorimeter cells with energy $E^{i}$ and longitudinal
energy $p^{i}_{Z}$.  To reduce uranium noise, the cell
energies were required to be greater
than 60 MeV (110 MeV) for the electromagnetic (hadronic) cells.
This calculation assumes that the scattered electron was not
detected in the calorimeter.  For DIS events in which the electron deposits
energy but is not identified in the calorimeter,
\yjb\ will be near unity.  Therefore, for further rejection of DIS events,
we required \yjb\ $< 0.7$.
To reject proton-gas interactions, \yjb\ $\ge 0.2$ was required.
These requirements correspond approximately to $0.2 < y < 0.85$.

\item To remove charged current background and cosmic ray showers,
a cut on $p_T\hspace{-4 mm}\slash$ $/\sqrt{E_T} < 1.5$ GeV$^{1/2}$
was imposed, where $p_T\hspace{-4 mm}\slash$ $ $ is the total transverse
momentum and $E_T$ is the total transverse energy of the event.

\end {itemize}

After these selection cuts, a sample of 99,894 events remained,
corresponding to an integrated luminosity of 0.55 pb$^{-1}$.
The estimated
proton-gas and DIS background contributions are 0.4\% and 1 to 2\%,
respectively.  Cosmic ray and electron-gas backgrounds are negligible.

\section{Monte Carlo simulation}

In the following, the data are compared to Monte Carlo simulations based on
the PYTHIA 5.6 \cite{pythia} event generator which includes
leading-order QCD calculations.  The HERWIG 5.7 \cite{herwig} event generator
was used to check the PYTHIA results.
The cut on the minimum transverse
momentum, $\hat p_{Tmin}$, of a hard scatter was set at 2.5 GeV.
In PYTHIA, the photon flux is calculated using the Weizs\" acker-Williams
approximation.
The parton densities used were GRV LO~\cite{grv}
for the photon and MRSD$_-$~\cite{MRSD} for the proton.
For comparison, we also used the
parameterization LAC1~\cite{LAC1} for the photon.

The generated events were passed through a detector simulation based on
GEANT 3.13~\cite{geant}.
The same reconstruction program that was used in the data analysis was
applied to the generated events.
The generated Monte Carlo event sample was obtained
by combining the resolved and direct samples in proportion to the
generated Monte Carlo cross sections (approximately 7:1 for the GRV
photon parton densities).

In PYTHIA the distribution
of the intrinsic transverse momentum, $k_t$, of the partons in the proton
and in the resolved photon
is parameterized by the distribution, $dN/dk_t^2 \propto
e^{-k_t^2/k_{0}^2}$, where $k_0$ is a parameter
which determines the hardness of the $k_t$ spectrum.  The default
value of $k_0$ for both the proton and the resolved photon
is 0.44 GeV.  An option in PYTHIA allows events to be
generated using a different functional form for the $k_t$
spectrum and a different value of $k_0$.  This option has been
used to generate events with a harder $k_t$ spectrum (see section 8).

\section{Analysis using the \Kt\ algorithm}

Previous analyses of photoproduction at HERA have implemented
a cone algorithm to find jets~\cite{res-ZE,res-H1,xg}.
This algorithm~\cite{cone} uses a cone of fixed radius
in pseudorapidity and azimuthal angle space and maximizes the
transverse energy within this cone.  It is therefore well
suited for high transverse energy jets.
The photon remnant, however, is expected to deposit energy
in the electron direction with low transverse energy.
Due to the rapid variation of pseudorapidity
in this region, and to the low transverse energy,
a cone algorithm cannot be used to identify the photon remnant.
Therefore, we chose to use the \Kt\ clustering algorithm~\cite{kt}.
The analysis was done in the laboratory frame instead of the
$\gamma p$ frame due to the uncertainties in the boost.

The \Kt\ algorithm
finds jets by iteratively merging clusters. Initially, clusters are
individual calorimeter cells.  For the Monte Carlo events,
the algorithm is also used to cluster generated particles (see section 7).
%generated particles are also used.
In the merging procedure,
the quantity \Kt\ is evaluated for each pair of clusters,
\[
k_{\perp} = 2E^2_{min}(1-{\rm cos}\theta_{nm}),
\]
where $\theta_{nm}$ is the angle between the two clusters $n$ and $m$,
and $E_{min}$ is the minimum value of the cluster energies $E_n$ and $E_m$.
In the small angle approximation, \Kt\ is the
transverse momentum squared of the lower energy cluster with respect
to the higher energy cluster.
A pseudo-particle with infinite momentum along the $Z$ axis is
included in the clustering
procedure to take the proton remnant into account.
The value of \Kt\ between the pseudo-particle and the other clusters
is calculated using the same formula as above.
When all of the \Kt\ values have been calculated, the two clusters with
the lowest \Kt\ value are merged.
The four-momentum of the new cluster is the sum of the four-momenta of the two
merged clusters.  The calculation of \Kt\ is
then repeated, replacing the two merged clusters with the new cluster.
The iteration continues until the energy-angle resolution
variable, Y (Y = \Kt/$E_T^2$, where $E_T$ is the total event energy transverse
to the beam axis),
becomes larger than some threshold, Y$_{cut}$.
The value of Y$_{cut}$ may be fixed or chosen on an event-by-event basis.
For high transverse energy jets, the \Kt\ cluster-finding algorithm
gives results which are similar to those obtained with the cone algorithm.

\section{\bf Photon remnant identification}

To identify the photon remnant, we begin with an intuitive
approach which is later justified using simulated
events.
As illustrated by Fig.\ 1b, resolved hard photoproduction events
have a final state which includes two high \Pt\ jets from the
hard scatter as well as photon and proton remnants.
%Therefore, we stop the merging when three clusters (in addition
%to the proton remnant) remain.  These
Since we expect to find three clusters (in addition to the proton remnant)
in each event, we choose the value of Y$_{cut}$ on an event-by-event basis
so that three clusters are found in each
event~\cite{old}.  These
three clusters then should correspond to the two clusters from the hard
scatter and the photon remnant.  Since the photon remnant is
expected to have low transverse momentum with respect to the beam
axis, the separation
between the photon remnant and the two jets from the hard scatter can be
achieved, to a first approximation, by associating the photon
remnant with the
cluster having the smallest transverse momentum.  For direct
events, where we do not expect to see a photon remnant, the lowest
transverse momentum cluster will either be part of the proton remnant
or part of one of the two high transverse momentum jets.  Therefore,
the pseudorapidity distributions of the lowest transverse momentum
clusters will be different for resolved and direct events.

In Fig.\ 2a-c, we show the (uncorrected) pseudorapidity
distributions of the three clusters obtained with the \Kt\ algorithm.
The data (full circles) and Monte Carlo events (histogram)
are shown normalized in the region $\eta^{cal} \le 1.6$.
While the two clusters with the highest \Pt$^{cal}$ (Fig.\ 2a,b) are
mostly found in the
$\eta^{cal} > 0$ region, the third cluster, with the lowest \Pt$^{cal}$,
(Fig.\ 2c) is observed mostly in the $\eta^{cal} < 0$ region,
{\it i.e.} in the photon direction.
Data and Monte Carlo expectations agree for the two highest \Pt$^{cal}$
clusters except in the forward region, $\eta^{cal} > 1.6$,
where we observe, in the data, an excess similar to that
already reported in our previous analyses \cite{jet-ze,dijet-ze}.
%This effect is not yet fully understood.
The peak observed in Fig.\ 2c, for the third cluster,
in the negative pseudorapidity region
is accounted for by the Monte Carlo
simulation including resolved and direct processes (full histogram).
The Monte Carlo distribution, however, is shifted slightly to lower values.
The direct process, which does not contain a photon
remnant, does not contribute to this
peak as shown by the dashed line in the figure.
Therefore, we are justified in using this method to separate direct
and resolved events.  In addition, the third
cluster, when it is in the
negative pseudorapidity region, can be associated with the photon remnant.

In order to maximize the possibility that the two highest \Pt$^{cal}$
clusters stem from the partons in the hard scatter and to minimize the
possibility that one of the two highest \Pt$^{cal}$ clusters is, in
fact, part of the proton remnant, we require
that the two highest \Pt$^{cal}$ clusters have high transverse
momentum (\Pt$_{1,2}^{cal} > 5$ GeV) and that they are well separated
from the forward region ($\eta_{1,2}^{cal} < 1.6$).

The distribution of the pseudorapidity of the third cluster, $\eta_3^{cal}$,
after the above cuts
and the requirement $E_3^{cal} > 2$ GeV, is shown in Fig.\ 2d.
The comparison with the distribution predicted for direct processes shows
that the events with $\eta_3^{cal} < -1$ are almost exclusively due to
resolved processes.  The agreement of the
Monte Carlo simulation for resolved plus direct contributions with the
data is not perfect; in the data there are somewhat fewer events with
large negative $\eta_3^{cal}$ values.
The difference between the data and the Monte Carlo simulation
is not improved when the photon parton parameterization LAC1 (dotted line)
is used instead of GRV LO.   For the following analysis, the
resolved event sample is selected by requiring $\eta_{3}^{cal} < -1$.

A measurement of the mean value of Y$_{cut}$ also demonstrates
the difference between the two event samples on either side of
$\eta_{3}^{cal} = -1$.
Higher values of Y$_{cut}$ indicate energetic clusters which are
spatially well separated.
For the events at high $\eta_3^{cal}$, the mean value of Y$_{cut}$ is 0.028.
For the low $\eta_3^{cal}$ events, the mean value of Y$_{cut}$ is 0.063.
This result suggests that the high
$\eta_3^{cal}$ events contain a significant number of direct, two-jet
events, for which the clustering procedure has been prematurely stopped.
For the sample at $\eta_{3}^{cal} < -1$, however, the separation
between the three clusters is quite distinct.

A total of 1370 events satisfy the cuts.
For these events,
the fraction of the photon momentum involved in the
hard scattering, $x_{\gamma}$, measured from the two highest $p_T^{cal}$
jets~\cite{xg}, peaks at low values as expected for resolved processes
(not shown).  By defining direct events, in this sample, as events with
$x_{\gamma} > 0.75$~\cite{dijet-ze}, we obtain a contribution from direct
processes of about 8\%.
The contribution from direct photon
interactions, estimated using LO direct Monte Carlo events, is also 8\%.

\section{\bf Correction procedure}

In the following, the data are corrected back to the hadron
level using the PYTHIA Monte Carlo program previously described.
For the Monte Carlo events, the \Kt\ algorithm is applied independently
at both the generated hadron level and the calorimeter cell level.
In both cases the
resulting clusters are sorted according to \Pt.
The detector level cuts in $\eta^{cal}$, \Pt$^{cal}$, $E_3^{cal}$, and
$y_{JB}$
correspond approximately to two hadron jets with \Pt$_{1,2} > $ 6.0 GeV and
$\eta_{1,2} < 1.6$, a remnant cluster with
$\eta_{3} < -1$ and $E_{3} > 2$ GeV, and $0.2 < y < 0.85$.
Therefore, the hadron level Monte Carlo
distributions were determined for events within this kinematic region.

The correspondence between hadron and calorimeter clusters
was determined by comparing the value of \Kt\ for
each pair of hadron and calorimeter clusters.  Each hadron
cluster was then matched with that calorimeter
cluster with which it had the lowest value of \Kt.
After all cuts were applied at both the hadron and calorimeter levels,
and neglecting the possible interchange of the labels, cluster 1 and
cluster 2, due to the sorting (this occurs in events in which the
two highest \Pt\ jets have similar \Pt),
all three clusters were found to be correctly matched for 97\% of the events.
%The two high-\Pt\ clusters were interchanged for 35\% of the events
For 3\% of the events, a high-\Pt\ cluster at one level was associated
with the proton remnant at the other level.
In less than 0.5\% of the events, the lowest \Pt\ clusters at the hadron
and calorimeter levels were not matched.

The experimental shifts and resolutions of the measured variables
were evaluated using the matched hadron and calorimeter clusters.
The average measured value of $\eta$
is shifted, with respect to the hadron level, by less than $-0.02(+0.08)$
and has a resolution of 0.07(0.43) units for the first two jets
(third cluster).
The measured transverse jet momentum, with respect to the beam, is reduced,
on average, by 15\% and has a resolution of 11\%.
This is due either to the magnetic field acting on low energy
particles or to dead material in the apparatus.
The value of $y$
is reconstructed with an average shift of $-0.14$ units and a
resolution of 10\%, also due to losses
in the beam pipe and detector effects.

On average, the third cluster
contains $75\pm20$\% of the photon remnant energy, as defined by the
third cluster energy at the hadron level.
Inactive material in front of the calorimeter results in an energy
loss of about 20\%,
independent of the cluster energy.  Particles lost in the beam pipe
account for the rest of the energy loss; this effect increases with
cluster energy, becoming comparable to the detector effects
at measured energies above 10 GeV.

The data were corrected with the following procedure.
First, the contamination from events outside the kinematic range was estimated
using Monte Carlo events and was subtracted bin-by-bin from
the measured distributions.
The resulting distributions were then corrected with
a correlation matrix that was generated using the matched hadron
and calorimeter clusters.  The unfolding procedure is
described in~\cite{giulio} and includes acceptance corrections for the
trigger and cuts described in section 3.
The statistical errors were estimated by randomly
varying, within their statistical errors, both the experimental
data (before the background subtraction)
and the Monte Carlo correlation matrix
and calculating the root mean squared
deviation in each bin.

The systematic uncertainties were estimated
by considering the following effects:
different photon parton densities,
a 5\% energy scale uncertainty in the Monte Carlo
simulation of the energy response of
the calorimeter, a variation of the cuts on the measured quantities
(\yjb, \Pt$_{1,2}^{cal}$, $\eta_{1,2}^{cal}$, $\eta_3^{cal}$), corrections
using the HERWIG Monte Carlo simulation, an increase in the
direct cross section by a factor of three relative to the resolved
one, and the use of calorimeter islands (see next section) instead of cells.
For the HERWIG systematic study, a fraction of the events was
generated with $\hat p_{Tmin}$
%the minimum transverse momentum of a hard
%scatter,
set to 5.0 GeV.
The systematic errors were evaluated by calculating the root mean squared
deviation between the original data point and each of the points found
using the systematic variations mentioned above,
weighting each deviation by its statistical errors.
In the following figures
the statistical errors are shown as the inner error bars; the outer bars
show the statistical and systematic uncertainties added in quadrature.

\section{\bf Photon remnant properties}

In Fig.\ 3, we show
the pseudorapidity, transverse momentum (with respect
to the $Z$ axis), and energy
distributions for the photon remnant, corrected back to the hadron level.
The average correction factors
are typically 1.2 and are approximately constant
for each of the three variables.

The $\eta_3$ distribution is shown in Fig.\ 3a.
The corrected data and the expectations from PYTHIA
(solid histogram) disagree in the negative $\eta_3$ region, as observed
in Fig.\ 2d.  The measured distribution peaks at higher values of $\eta_3$
than the Monte Carlo prediction.
A similar effect can be observed in the transverse momentum
distribution
(Fig.\ 3b), where the requirement $\eta_3 < -1$ is applied for this
and all following figures.
Here also, the data show a higher average value.
The distribution peaks at 1.5 GeV with a tail extending to 6 GeV.
The mean value of the photon remnant \Pt\ at the hadron level is
measured to be $\langle p_{T3} \rangle = 2.1 \pm 0.2$ GeV.
The Monte Carlo expectation is $\langle p_{T3} \rangle = 1.44 \pm 0.02$ GeV.

The energy distribution (Fig.\ 3c) peaks around 7 GeV
and extends to 20 GeV.
The solid histogram shows the Monte Carlo expectation
which agrees with the data,
except in the lowest bin where
the Monte Carlo expectation is about 30\% higher.

The unlikely possibility has been studied that an incorrect Monte
Carlo description of
the energy response in the RCAL might result in
the observed discrepancy between the data and the Monte Carlo expectation.
The RCAL energy scale was reduced by 10\% within 10$^\circ$
of the RCAL beam pipe to simulate additional inactive material.  The
discrepancy in the \Pt$_3$ distribution was not improved, and the
disagreement with the data for the $E_3$ distribution became worse.

The observation that the average transverse momenta of the
photon remnant is higher than
expected from PYTHIA,
is in qualitative agreement with the theoretical predictions
mentioned in the introduction.  Therefore,
following~\cite{drees}, we have compared the
data with Monte Carlo events generated
with a harder intrinsic transverse
momentum spectrum for the partons in the photon, {\it i.e.}
$dN/dk_t^2 \propto 1/(k_t^2 + k_{0}^2)$.
The ``PYTHIA high $k_t$'' results are shown as the
dotted histograms in Fig.\ 3.  The parameter
$k_0$ is determined by minimizing the $\chi^2$ between the Monte Carlo
hadron level and the corrected data \Pt$_3$ distributions.
The result is $k_{0} = 0.66 \pm 0.22$ GeV.  This corresponds to
$\langle k_t \rangle \approx 1.7$ GeV, as compared to 0.4 GeV for PYTHIA
with default parameters.
The agreement with the energy distribution is unchanged, while it is
considerably improved for
the $\eta_3$ and \Pt$_3$ distributions.
On the other hand, the distributions for the first two
clusters remain unchanged (not shown).
Equally good agreement between the data and the Monte Carlo simulation
has been achieved by reweighting the Monte Carlo events to the default
intrinsic $k_t$ spectrum (not shown).  In this case,
$k_0 = 1.90 \pm 0.21$ GeV, much higher than the default value of
$k_0 = 0.44$ GeV assumed in PYTHIA.
These results show that adjusting the intrinsic transverse momenta of the
partons in the photon is a way to improve the agreement between the data
and the Monte Carlo predictions.

One of the well-known properties of jets is that the average energy
transverse to the jet axis is limited as the jet energy increases.
In general, this results in an average energy transverse to
the jet axis per particle, $\langle E_T^i\rangle$, of the order
of a few hundred MeV\@.
In measuring this quantity for the photon remnant, we used
calorimeter islands.  Islands are groups of calorimeter cells
more closely related to particles than individual calorimeter cells
and therefore represent a better choice for this measurement.
The analysis performed with calorimeter cells provides an upper limit
on the systematic uncertainties of this measurement.
Figure 4a shows the average value of $\langle E_T^i\rangle$ versus the
cluster energy
for the third cluster, both for the data after corrections and for
the Monte Carlo simulation.
The corrections include the effect of particles lost down the beam pipe.
The largest component of the systematic errors (+30\%) comes
from using calorimeter cells instead of islands.

The mean value of $\langle E_T^i\rangle$
starts below 200 MeV, and slowly increases with the remnant energy, $E_3$.
The total cluster energy, on the other hand, spans a range from 3 to 21 GeV.
This result demonstrates that the photon remnant exhibits limited
transverse energy per particle, as expected for jet-like objects.
Further support to this conclusion is given in Fig.\ 4b, which shows
the average values of the corrected total transverse
($\Sigma_i E_T^i$) and total longitudinal ($\Sigma_i E_L^i$) energy
of the third cluster, with respect to the cluster axis,
as a function of the energy of the cluster.
The longitudinal component increases much faster than the transverse
energy, and most of the cluster energy is along the cluster axis.
This is consistent with a jet-like structure of the photon remnant.

We also studied how
the energy is distributed around the axis of the third cluster.
Figure 4c shows the corrected energy flow of the third
cluster as a function of $1-\rm{cos}\Theta$ for both the
data and the Monte Carlo simulation.
Here, $\Theta$ is the angle of the particle with respect to
the cluster axis.  This is effectively a plot of the energy deposited in rings
of fixed area centred on the cluster axis.
Because there is not a simple correspondence between particles and
calorimeter cells, it is difficult to construct
a correlation matrix between the generated
(hadron energy) and experimental (calorimeter cell energy) quantities.
Therefore, these distributions are corrected bin-by-bin.
The average correction factor is around 1.15.
In Fig.\ 4c, we have required reconstructed (uncorrected)
jet energies between 8 and 14 GeV and hadron level jet energies between
8 and 15 GeV.
The statistical errors are the error on the mean.
The energy distribution for the data is quite collimated.
The Monte Carlo simulation agrees very well with the data, indicating
that the fragmentation of the remnant is understood.

\section{\bf Comparison of the photon remnant with jets from hard scattering}

Having established the jet-like properties of the photon remnant, we
next compare it with the jets originating
from parton hard scattering.  The comparison of these two types of jets
is of interest because
one is the debris of the photon and is a low-\Pt\ jet, with \Pt\ typically
well below 6 GeV, while the other two jets come from the hard
scattering of the
partons in the photon and proton and are high-\Pt\ jets, with a minimum \Pt\
of 6 GeV.
A comparison between low-\Pt\ and high-\Pt\
jets has been proposed as a method of studying the hadronization
process~\cite{isrpp}.

Figures 4d, e, and f present the results
of this comparison for the average value of $\langle E_T^i\rangle$,
$\langle \Sigma_i E_T^i \rangle$ and
$\langle \Sigma_i E_L^i \rangle$,
and $1-\rm{cos}\Theta$, respectively.
In Fig.\ 4f, we again require reconstructed (uncorrected)
jet energies between 8 and 14 GeV, and hadron level jet energies between
8 and 15 GeV.  This cut is especially important for this figure in order
to compare jets with approximately equal energies.
In all figures, good agreement between the third cluster and the two
hard jets is observed.
{}From these comparisons we conclude that,
in the kinematic region and for the variables studied,
the low-\Pt\, photon remnant jet exhibits the same
hadronization characteristics as the
high-\Pt\ jets originating from the hard scattering process.

\section{\bf Conclusions}

For the first time, in a sample of quasi-real photon-proton collisions,
the photon remnant produced in resolved photon interactions
has been isolated.  The selected events contain
two high-\Pt\ jets with \Pt\ $> 6$ GeV
and $\eta < 1.6$, and $130 < W_{\gamma p} < 270$ GeV.
The properties of the photon remnant, as defined by a cluster with
$\eta_3 < -1$ and $E_3 > 2$ GeV, are
studied and shown to exhibit a collimated energy
flow with a limited transverse
energy with respect to the cluster axis, characteristic of a
jet structure.

The leading order QCD Monte Carlo simulation, with
default parameters,
%A model of resolved photon production in LO QCD
does not reproduce the pseudorapidity distribution or the transverse momentum
distribution (with respect to the incident photon) of the photon remnant.
The mean value of \Pt\ for the photon remnant, $2.1 \pm 0.2$ GeV,
is substantially larger than the Monte Carlo expectation.
Better agreement can be obtained by increasing the average
intrinsic transverse momenta of the partons in the photon to about
1.7 GeV.  These results are in qualitative agreement with
theoretical expectations of substantial mean transverse momenta for the
photon remnant.

The photon remnant has also been compared, in the laboratory frame,
with the two high-\Pt\
jets originating from the parton hard scattering.
Although the origins of these two types of jets may be quite different,
within the present statistics and in the kinematic range studied,
they exhibit similar properties for the energy flow and the transverse and
longitudinal energy with respect to the jet axis.

\section{\bf Acknowledgements}
We thank the DESY directorate for their strong support and encouragement.
The remarkable achievements of the HERA machine group were essential for the
successful completion of this work, and are gratefully appreciated.
We also gratefully acknowledge the support of the DESY computing
and network services.

\parskip 0mm
\begin{figure}[htbp] %figure 1
\epsfxsize=6 in
\epsfysize=6 in
\epsfbox[-30 0 537 567]{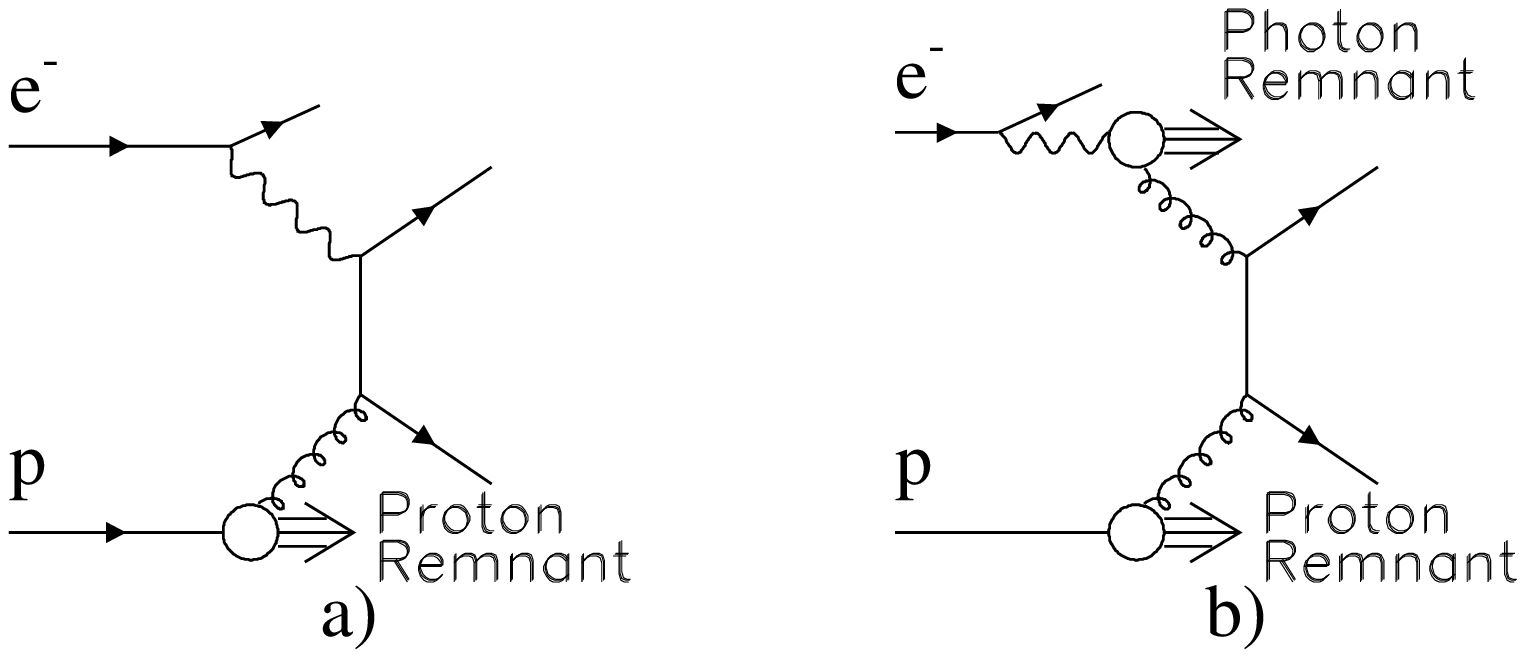 }
\vspace{0.4 in}
\caption[]{\label{f:fig0} Examples of leading order diagrams
for (a) direct and (b) resolved
hard photoproduction interactions.}
\end{figure}

\parskip 0mm
\begin{figure}[htbp] %figure 2
\epsfxsize=6 in
\epsfysize=6 in
\epsfbox[-30 0 537 567]{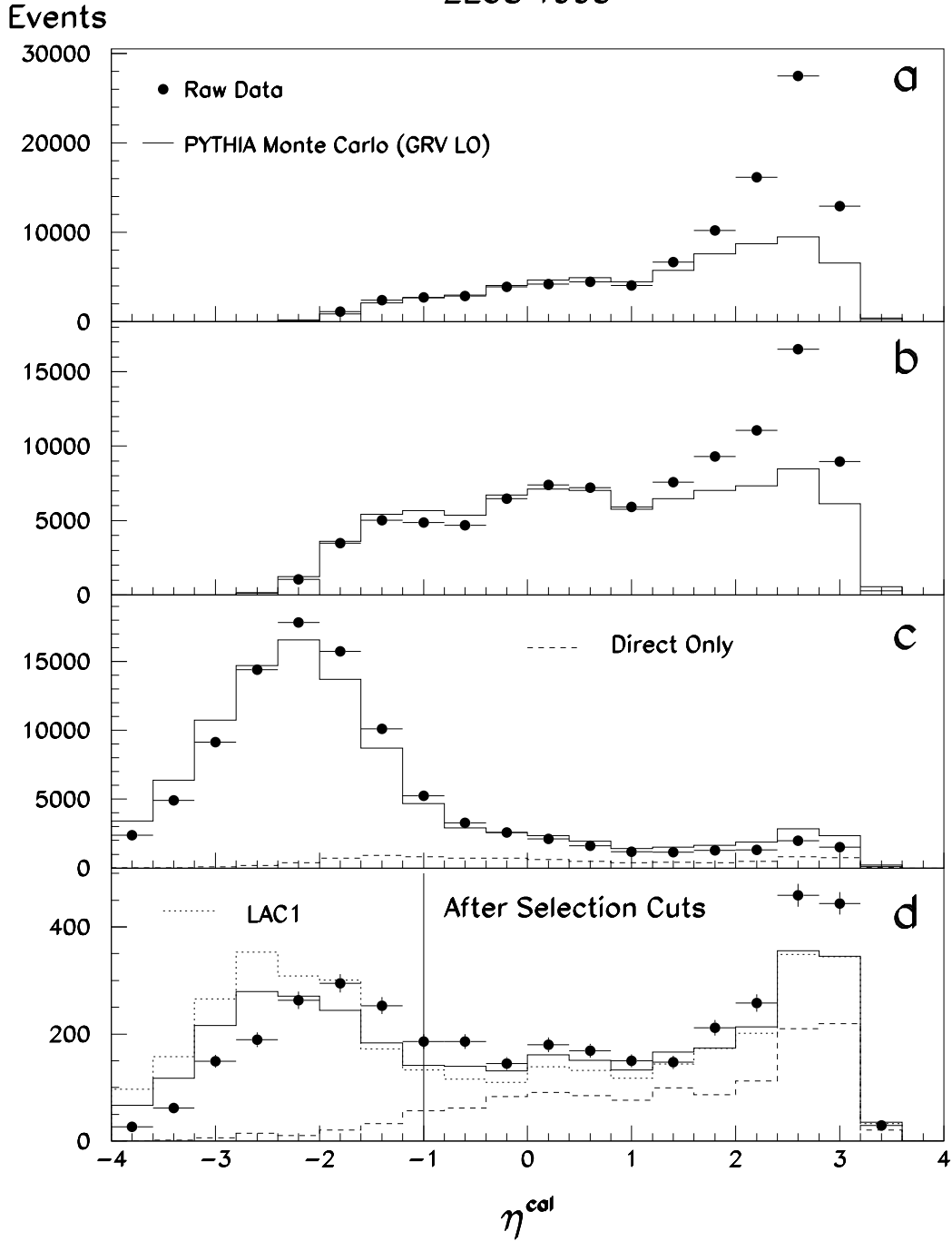 }
\vspace{0.4 in}
\caption[]{\label{f:fig1} Pseudorapidity distributions of clusters
for the inclusive photoproduction sample.  The three clusters
are sorted by \Pt$^{cal}$
with (a) having the highest, (b) the second highest, and (c) the lowest
\Pt$^{cal}$.
Each Monte Carlo distribution is independently normalized to
the data in the region $\eta^{cal} < 1.6$.  In (c) and (d) the
direct contribution alone is shown as the dashed line.
The $\eta^{cal}$ distribution of the lowest \Pt$^{cal}$ cluster
(as in (c)) is
shown again in (d) after requiring \Pt$^{cal}_{1,2} > 5$ GeV,
$\eta^{cal}_{1,2} < 1.6$, and
$E_3^{cal} > 2$ GeV.  Resolved events are selected by requiring
$\eta_3^{cal} < -1$ (vertical line in (d)).  The dotted line shows the
expectation using the LAC1 photon parton parameterization. }
\end{figure}

\parskip 0mm
\begin{figure}[htbp] %figure 3
\epsfxsize=6 in
\epsfysize=6 in
\epsfbox[-30 0 537 567]{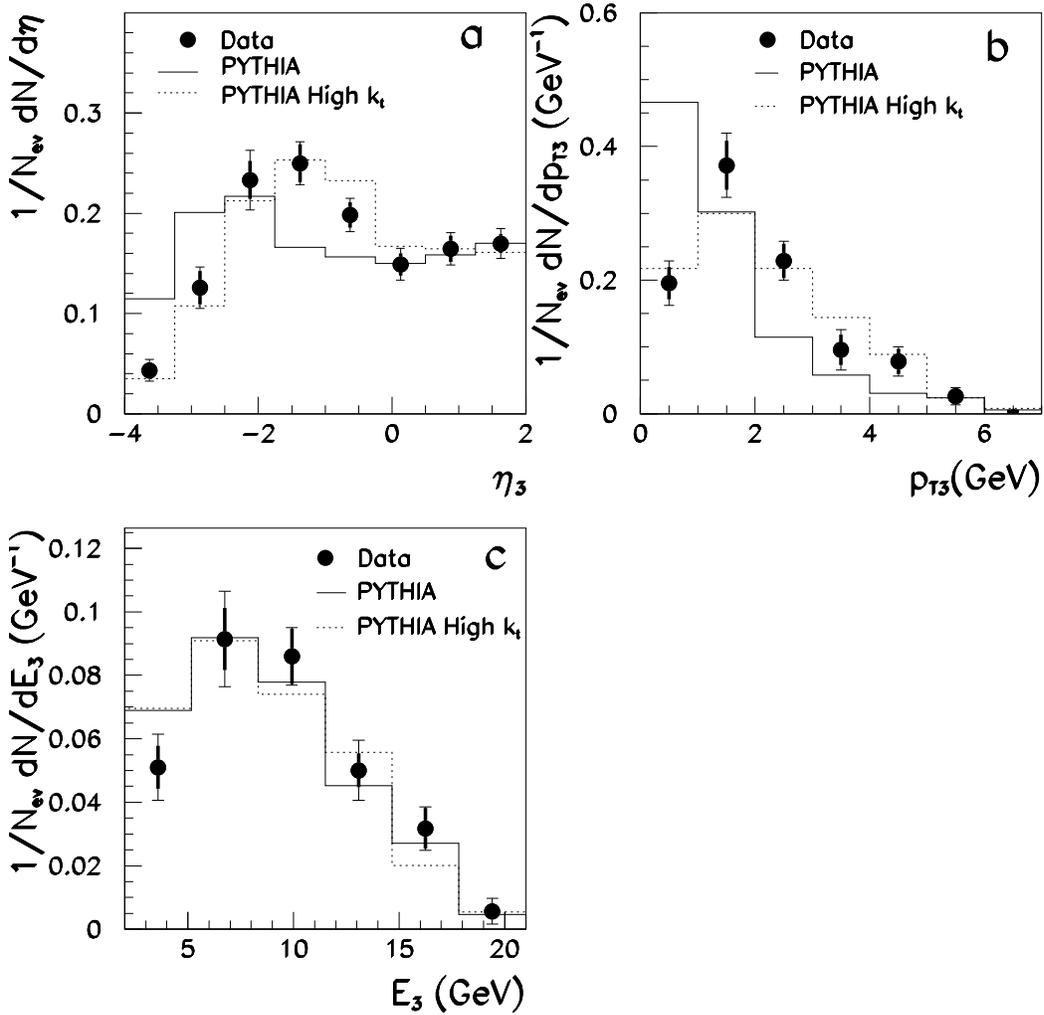 }
\vspace{0.4 in}
\caption[]{\label{f:fig2}
(a) Pseudorapidity distribution of the third cluster corrected back to the
hadron level. (b) Corrected
\Pt\ distribution.  (c)
Corrected energy distribution.  In (b) and (c) we require
$\eta_3 < -1$.
The solid histograms are
the hadron level distributions given by the default version of PYTHIA
(Gaussian with $k_0 = 0.44$ GeV).
For each of the figures, the dotted line shows
the Monte Carlo predictions with
$dN/dk_t^2 \propto 1/(k_t^2 + k_{0}^2)$ and $k_0 = 0.66$ GeV,
corresponding to $\langle k_t\rangle \approx 1.7$ GeV.}
\end{figure}

\parskip 0mm
\begin{figure}[htbp] %figure 4
\epsfxsize=6 in
\epsfysize=6 in
\epsfbox[-30 0 537 567]{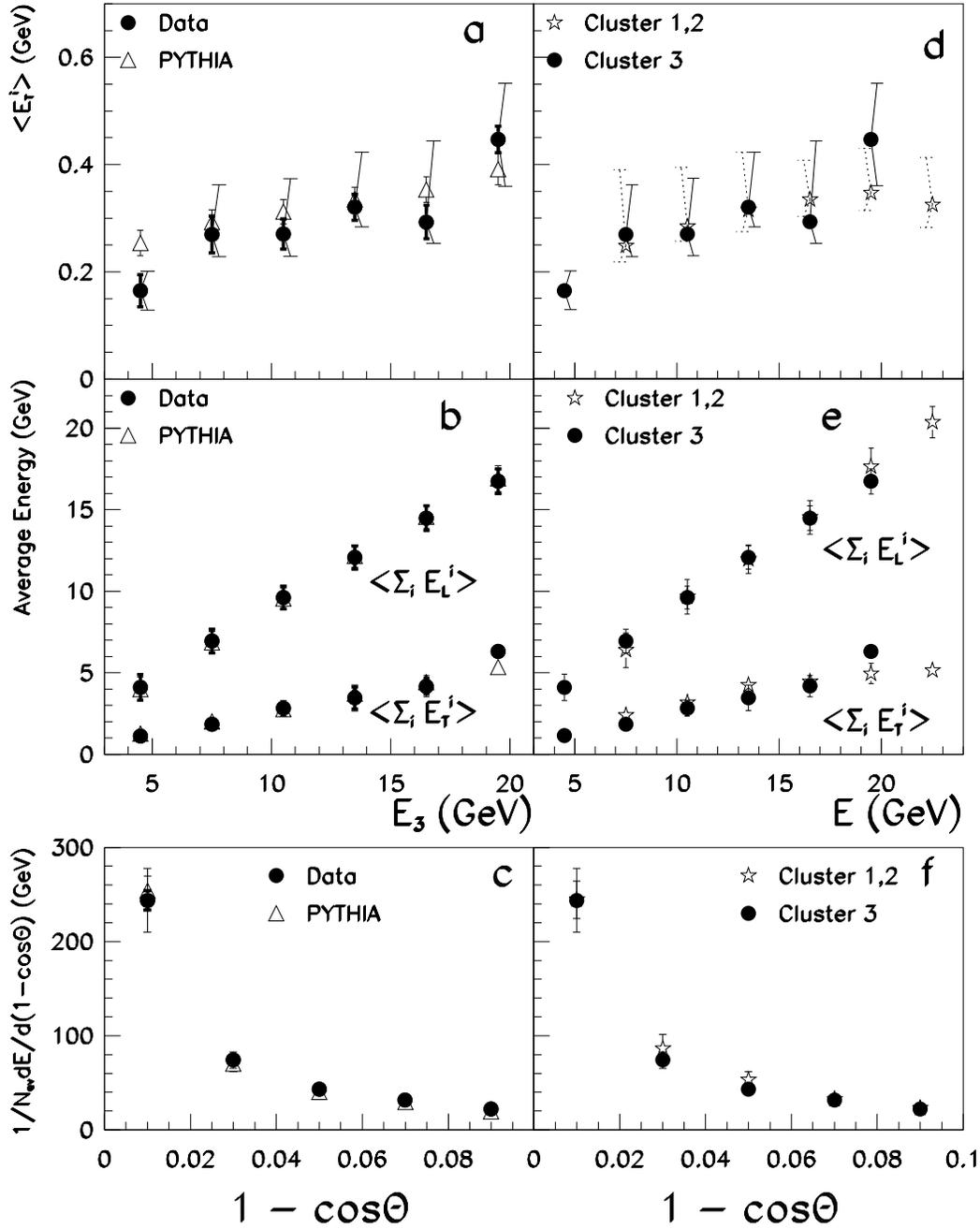 }
\vspace{0.4 in}
\caption[]{\label{f:fig3} (a-c) Comparison between the data and
the hadron level Monte Carlo expectations.  (a) The mean value of
$\langle E_T^i \rangle$, the average energy transverse to the cluster
axis per particle, as a function of the cluster energy.
(b) The average values of the
total transverse ($\Sigma_i E_T^i$) and total longitudinal
($\Sigma_i E_L^i$) energy. (c) The flow of energy around the cluster
axis. (d-f) Comparison between the photon remnant (cluster 3)
and the two high-\Pt\ jets.
The error bars show the systematic and statistical errors added in
quadrature.}
\end{figure}

\end{document}